\documentclass[aps,prb,showpacs,superscriptaddress,twocolumn]{revtex4-1}

\usepackage{amsmath,graphics}
\usepackage[next]{inputenc}
\usepackage[dvips]{epsfig}
\usepackage[colorlinks=true,citecolor=blue,linkcolor=blue]{hyperref}
\usepackage{bbm}
\usepackage{bbold}
\usepackage{booktabs}
\usepackage{multirow}
\usepackage{hhline}
\usepackage{graphicx}
\usepackage{amsmath}
\usepackage{multirow}

\def\be{\begin{equation}}
\def\ee{\end{equation}}

\def\bi{\begin{itemize}}
\def\ei{\end{itemize}}
\def\bn{\begin{enumerate}}
\def\en{\end{enumerate}}
\def\bea{\begin{eqnarray}}
\def\eea{\end{eqnarray}}

\def\ba{\begin{array}}
\def\ea{\end{array}}
\def\bd{\begin{displaymath}}
\def\ed{\end{displaymath}}

\begin{document}

\title{Spin-Orbit Interaction and Kondo Scattering at the PrAlO$_3$/SrTiO$_3$  Interface: Effects of Oxygen Content}

\author{Shirin Mozaffari}
\affiliation{Department of Physics, The University of Texas at Austin, Austin, TX 78712, USA}

\author{Samaresh Guchhait}
\affiliation{Microelectronics Research Center, The University of Texas at Austin, Austin, TX 78758, USA}

\author{John T. Markert}

\email{markert@physics.utexas.edu}

\affiliation{Department of Physics, The University of Texas at Austin, Austin, TX 78712, USA}

\begin{abstract}
 We report the effect of oxygen pressure during growth ($P_{O_{2}}$) on the electronic and magnetic properties of PrAlO$_3$ films grown on  $\rm TiO_{2}$-terminated  SrTiO$_3$ substrates. Resistivity measurements show an increase in the sheet resistance as $P_{O_{2}}$ is increased. The temperature dependence of the sheet resistance at low temperatures is consistent with Kondo theory for $P_{O_{2}} 	\ge 10^{-5}$ torr. Hall effect data exhibit a complex temperature dependence that suggests a compensated carrier density.   We observe behavior consistent with two different types of carriers at interfaces grown at $P_{O_{2}}	\ge 10^{-4}$ torr.  For these interfaces, we measured a moderate positive magnetoresistance (MR) due to a strong spin-orbit (SO) interaction at low magnetic fields that evolves into a larger negative MR at high fields. Positive high MR values are associated with samples where a fraction of carriers are derived from oxygen vacancies. Analysis of the MR data permitted the extraction of the SO interaction critical field ( e.g. $ H_{SO}=$1.25 T for $P_{O_{2}}=10^{-5}$ torr). The weak anti-localization effect due to a strong SO interaction becomes smaller for higher $P_{O_{2}}$ grown samples, where MR values are dominated by the Kondo effect, particularly at high magnetic fields. 
                
\end{abstract}
\date{\today}

\maketitle
\section{Introduction \label{intro}}
The broken translational symmetry at an interface between complex metal oxides gives rise to a number of remarkable phenomena with rich physics often due to charge, spin, orbital, or atomic reconstruction, such as superconductivity \cite{SCthinfilms}, metal-insulator transitions \cite{MIT,mehdi}, and giant magnetoresistance \cite{GMR}. These phenomena can be entirely different from those innate to the constituent materials\cite{Hwang2012,AR}. A recent example of such interface phenomena is the two-dimensional electron gas (2DEG) at the interface of LaAlO$_3$ (LAO) and $\rm TiO_{2}$-terminated  SrTiO$_3$(STO) \cite{ohtomo}. Despite that LAO and STO are both band insulators and non-magnetic, their interface is found to be conductive \cite{ohtomo}, superconductive \cite{Reyren}, ferromagnetic \cite{Magnetism,magnetic} and exhibits an electric-field tunable metal-insulator transition \cite{Caviglia}, which all are the result of electronic reconstruction at the interface. Most of the mentioned properties can be altered by applying a back gate, which varies the electron density at the interface\cite{Caviglia}. It is generally agreed that the electron gas is formed by transfer of electrons from the polar layers of LAO to the top TiO$_2$ layer of STO \cite{sharp}.  In addition to the electronic reconstruction at the interface, doping the interface by interdiffusion of La/Sr atoms \cite{interdiffusion} and by the presence of oxygen vacancies that are created in the STO substrate during thin film deposition \cite{Ovacancy} contributes to the conductivity. In the electronic reconstruction model, donors ultimately originate from the surface of LAO. They come from the interface or within STO in the other two  mechanisms. The reason for cation mixing is thought to be the reduction in the dipole energy at the interface, and it has been shown that La$_{1-x}$Sr$_x$TiO$_3$ (${0.05\textless x \textless 0.95}$) is metallic\cite{interdiffusion,La_Sr_diffusion2}. It was also experimentally\cite{Ovacancy2,Ovacancy3} and theoretically\cite{vacancy-theory,vacancy-theory2} demonstrated that varying oxygen pressure during growth results in dramatic changes in interfacial conductivity. Moreover, conductivity found at the interface of STO and amorphous LAO\cite{LAO-amorph,LAO-amorph2}, suggest that oxygen vacancies can be dominant source of mobile carriers. Oxygen post-annealing of the amorphous samples fills the oxygen vacancies, and the interface becomes insulating.\par
Strong spin-orbit (SO) coupling effects whose magnitude can be modulated by the application of an external electric field has been observed for the LAO/STO interface through magnetotransport studies\cite{LAO-Rashba}. This SO coupling effect that arises from the broken inversion symmetry at the interface lifts the spin degeneracy of the 2DEG and is known as Rashba spin splitting\cite{SO-book}. Unlike in semiconductor heterostructures, interfacial carriers at the LAO/STO interface reside in the 3\textit{d}-\textit{t$_{2g }$} conduction band of STO, potentially affecting the strength of the SO coupling\cite{SO-theory}. Density-functional-theory calculations\cite{SO-theory} show a large SO coupling effect around the $ \Gamma $ point, either with standard \textit{k}-linear Rashba spin splitting or with a cubic dependence on the wave vector, depending on the sign of the \textit{xy} sub-band splitting.  
 \par
Similar interfaces such as the interface of STO and PrAlO$_3$\cite{Pr}, NdAlO$_3$ \cite{Nd}, LaGaO$_3$\cite{LGO}, NdGaO$_3$\cite{NdGaO} and LaTiO$_3$\cite{LTO}, where the film is polar, have all shown conductivity. Most of these interfaces have a higher lattice mismatch and stronger localization of carriers relative to that of LAO/STO, which result in different transport mechanisms and properties. \par
In order to examine the importance of the rare-earth cation R on the transport properties of the 2DEG at RAlO$_3$/STO interfaces, we have investigated the system with R=Pr, that is the interface between PrAlO$_3$ (PAO) and STO. Among the aluminates, PAO has the closest lattice constant to LAO (due to the multivalancy of Ce, it is hard to make stable thin films of CeAlO$_3$), and thus this helps to lessen the effect of strain as much as possible. The effective magnetic moment of Pr is 3.6 Bohr magnetons, due to \textrm{Pr$^{3+}$} \textit{f} electron states, often hybridized, while closed shell \textrm{La$^{3+}$} is diamagnetic\cite{SS-Book}, thus this rare-earth substitution can lead to magnetic properties in an aluminate structure. In bulk, PAO is known to be paramagnetic and has mostly been studied for its structural transitions below room temperature which are absent in LAO\cite{Pr-magnetic}. 
\par
This paper is organized as follows. Section \ref{Experiments} contains our experimental method for synthesizing and characterizing samples. Section \ref{Results} contains our experimental results of resistance as a function of temperature and magnetoresistance as a function of magnetic field for samples that are made at different oxygen pressures during growth. Comparison with theoretical concepts are made. A summary and conclusions are given in Section \ref{Conclusion}.

\section{Experiments}\label{Experiments}
Epitaxial thin films were grown in a custom-built Pulsed Laser Deposition (PLD) vacuum chamber using a Lambda Physik COMPex 201 KrF (248 nm) pulsed excimer laser. The  TiO$_2$-terminated SrTiO$_3$ (001) substrate was prepared by etching the substrate in buffered hydrofluoric (HF) acid for one minute and then by thermal annealling at $\mathrm{950 ^{\circ}C}$ for 2 hours. To keep the effect of the substrate defect constant, a single substrate was cut into equal pieces to be used for depositing the different samples. PAO epitaxial thin films were prepared in the oxygen partial pressure ($P_{O_{2}}$) range of $6\times10^{-6}$ -- $1\times10^{-3}$ torr at $\mathrm{750 ^{\circ}C}$. The laser pulse energy was 100 mJ, the fluence of the laser was $\sim$  3 \textrm{J{cm$^{-2}$}} and the repetition rate was 1 Hz. After deposition, the samples were post annealed for 30 minutes and then cooled down to room temperature, both in the same oxygen pressure as during deposition. The quality of the terminated substrate and the thin film samples were further confirmed by atomic force microscopy (AFM) using a Veeco Nanoscope, ex situ at room temperature using tapping mode. The crystal phase and epitaxial growth of the films were later confirmed by x-ray diffraction (XRD) with a Philips X'PERT diffractometer using CuK$\alpha$ x-rays ($\lambda$ = 1.54056 \r{A}). Alignment was achieved using the (002) reflection of STO. The buried interface was contacted in a Van der Pauw geometry by ultrasonic welding using a West Bond Inc. model 7476D manual aluminum wedge-wedge wire-bonder. Transport properties were measured using a 9 T Quantum Design physical property measurement system (PPMS).  The growth oxygen pressures reported were from an ion gauge in close proximity to the substrate, but should be considered approximate, since pressure gradients are inherent between oxygen source and vacuum pump.

\section{Results and Discussions}\label{Results}
\subsection{Crystal structure and surface analysis}
\begin{figure}[h]
	\includegraphics[width=\linewidth, trim=0 5cm 0 0cm, clip]{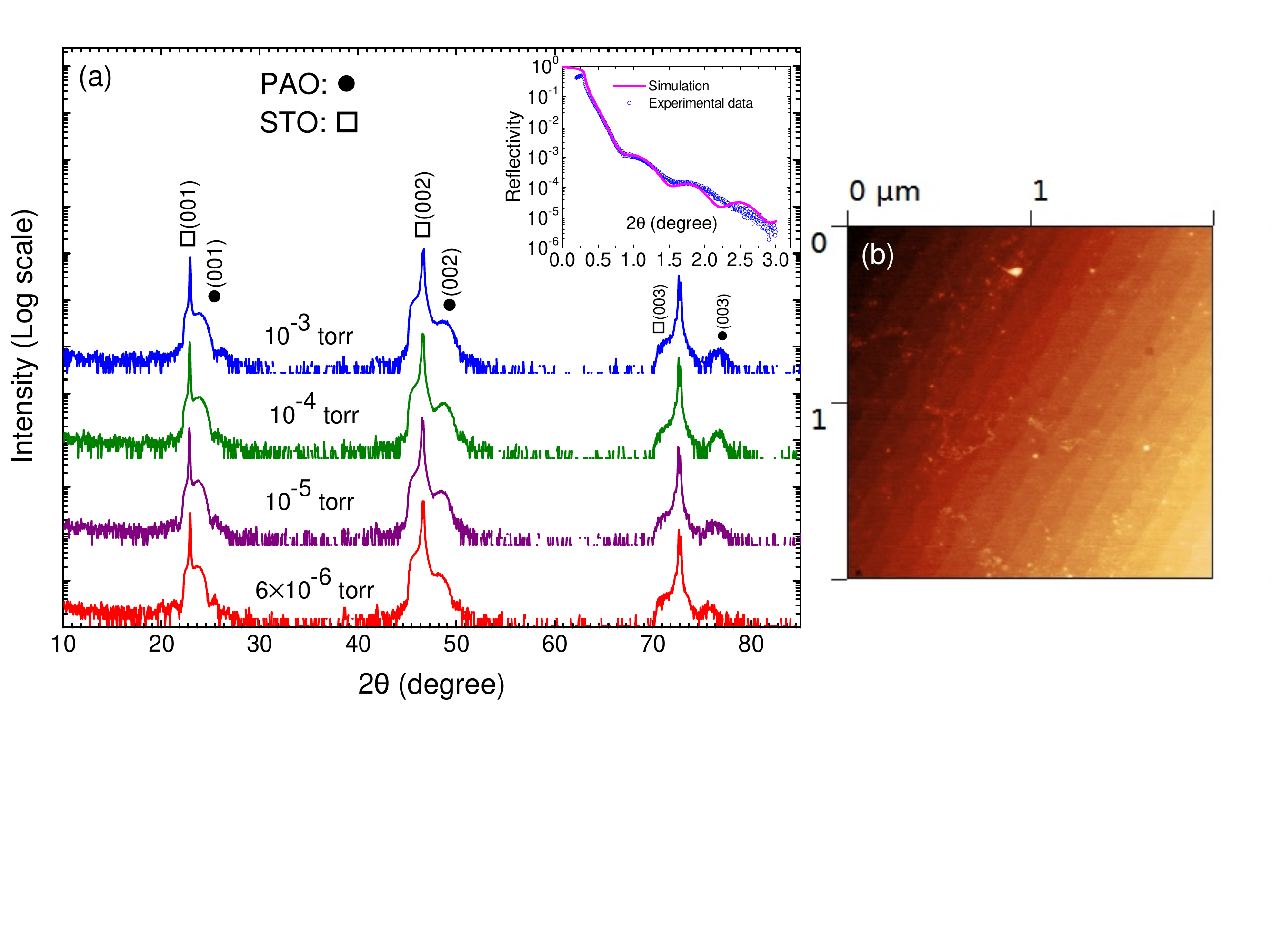}
	\caption{(Color online) (a) XRD data fro PAO/STO samples grown at different oxygen pressures. The XRR pattern of the $10^{-3}$ torr sample is shown in the inset (blue symbols). The pink line is a simulation. (b) AFM topography of PAO/STO grown at $1\times 10^{-3}$ torr.} \label{XRD}
\end{figure}

Figure \ref{XRD} (a) shows an XRD scan of epitaxial PAO films on STO (001) that were grown at different $P_{O_{2}}$. The thin films were grown along \textit{c}-axis since only the (00\textit{l}) reflections of the PAO are present. The inset shows an x-ray reflectivity (XRR) scan of the $1\times10^{-3} $ torr sample. From a simulation fit, the thickness is estimated to be 6.2 nm, or 15.5 unit cells (uc). Similar results were obtained for the other samples. There was no significant change in the structure, crystallinity, and quality of these samples as a function of oxygen pressure during growth. Figure \ref{XRD} (b) depicts the AFM topography of PAO/STO sample grown at $1\times10^{-3}$ torr. The step terrace structure of the substrate is maintained, verifying the layer-by-layer step-flow growth. \par
\subsection{Sheet resistance and Kondo effect}
\begin{figure}[hb]
	\includegraphics[width=9 cm]{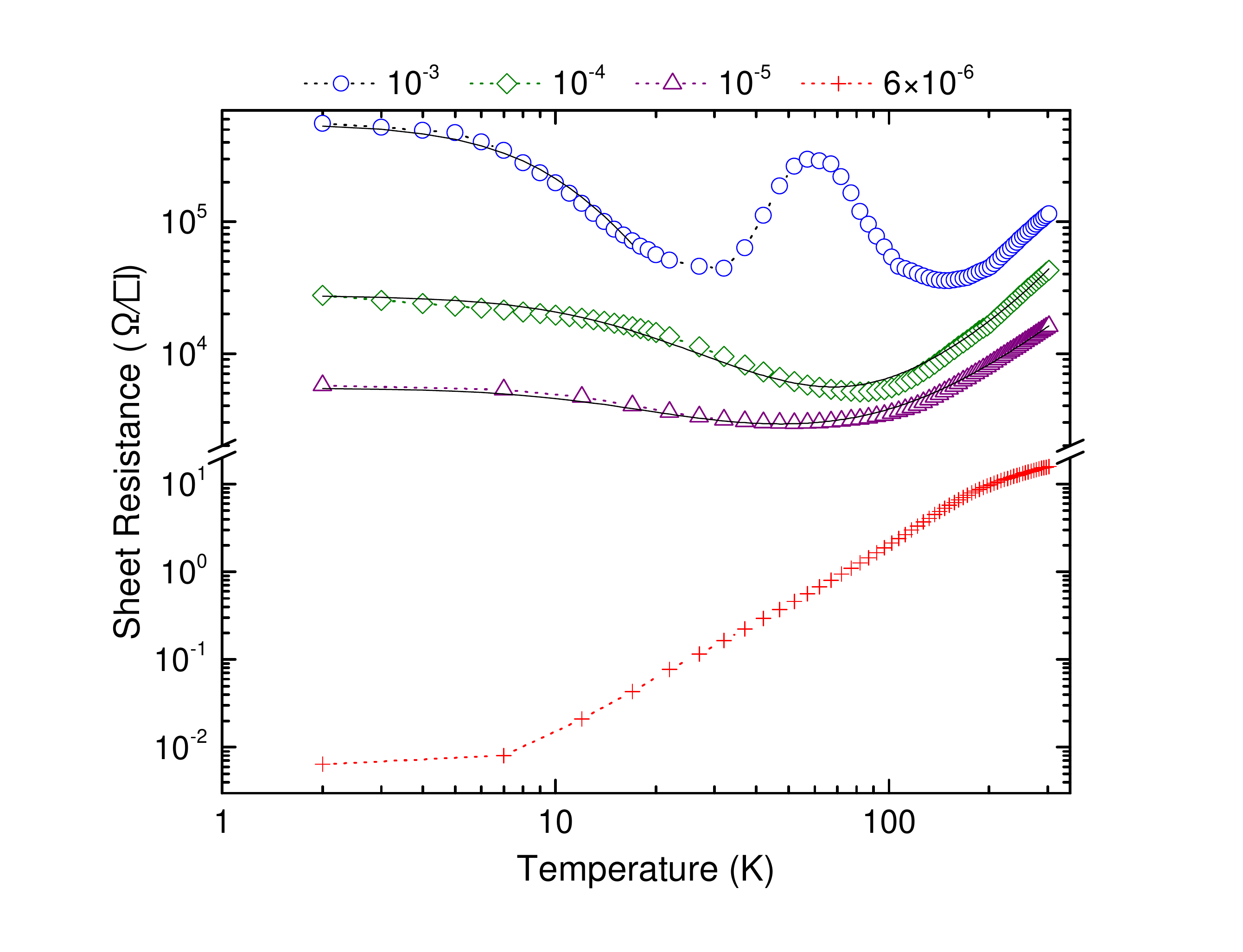}
	\caption{(Color online) Sheet resistance of PAO/STO interface synthesized at different oxygen pressure as a function of temperature. Solid curves are a fit using Eq.\ref{R(T)-eq} and \ref{fit-eq}. } \label{RvsT}
\end{figure}

In Figure \ref{RvsT} we present the dependence on $P_{O_{2}}$ of the sheet resistance $\textit{R$_s$}$ of the interfaces of PAO and STO. We observe that the samples become less conductive for higher $P_{O_{2}}$. The room temperature resistivity increases as the samples are prepared at higher
 oxygen pressure. While there is only less than a factor of 2 in growth pressure difference between the $6\times10^{-6}$ torr and $1\times10^{-5}$ torr samples, their room temperature resistivities differ by three orders of magnitudes.  In the case of the LAO/STO interface, it is known that oxygen vacancies can provide intragap donor levels close to the conduction band of STO\cite{AdvMat}. Higher oxygen pressure during growth tends to fill the oxygen vacancies, thus causing the room temperature $\textit{R$_s$}$ to increase. The lowest oxygen-pressure-grown sample shows a metallic behavior from room temperature down to 2 K. The resistivity for this sample levels off below 10 K. We also treated an STO substrate under the deposition conditions, without an actual film on it, but it turned out to be very resistive and we were not able to measure any measurable conductance for it. The question of whether oxygen vacancies or cation intermixing are in fact responsible for the observed conduction still remains. The very low sheet resistance indicates that the conduction layer is much thicker for $6\times10^{-6}$ torr sample and the transport must occur in a region many layers thick.
For $10^{-5}$ and $10^{-4}$ torr O$_2$ growth pressure, the resistivity of the PAO/STO interface has a metallic trend from room temperature down to 50 K and 75 K, respectively, and nonmetallic below those temperatures. For the $10^{-3}$ torr sample, the minimum of $\textit{R$_s$}$ at \textit{T}$_m$  happens at a higher temperature of 150 K, and it is accompanied by an unusual maximum at around 57 K. Currently, we do not have a clear explanation on the unusual maximum a around 57 K which then is followed by a minimum at 32 K. We note that the temperature of the maximum coincides with the melting point of oxygen and may be related to variation in  trapped oxygen mobility.\par
Reference \onlinecite{Pr} has examined interfaces of PrAlO$_3$ and NdAlO$_3$ (NAO) with SrTiO$_3$ at $P_{O_{2}}$ = $10^{-3}$ torr. Our observed maximum at 57 K in $\textit{R$_s$}$ is absent in their data. This may be due in part to different absolute growth pressures.\par
 
At very low temperatures, all the samples, except the lowest $P_{O_{2}}$ one, exhibit insulating behavior by an upturn in $\textit{R$_s$}$. The upturn is substantially stronger for the high $P_{O_{2}}$ sample. 
The upturn in $\textit{R$_s$}$ at low temperatures has been seen in some of the reported results for the interface of LaAlO$_3$/SrTiO$_3$ that were grown at pressures higher than $1\times10^{-4}$ torr, or in samples thicker than 15 uc\cite{LAO-15uc}. These variations in $\textit{R$_s$}$ suggest that different types of transport mechanisms are operating at the interface. Some of the scattering processes that result in an upturn in resistivity are: scattering from localized magnetic moments (Kondo effect), weak localization (WL), and variable range hopping due to strong localization. The presence of a saturating resistance below \textit{T}$_m$ is the signature feature of the Kondo effect and our data are best fit to a Kondo model. In the Kondo model an impurity in a metal couples to reservoir of itinerant electrons by an antiferomagnetic exchange coupling forming a virtual bound state. This coupling can result in a contribution to the resistivity through a universal function of $ R_K(T/T_K) $, in units of Kondo temperature ($\textit{T$_K$}$). \par
Metallic resistance including a Kondo effect term is given by:
\begin{equation}\label{R(T)-eq}
R^{fit}\left( T\right) =R_{0}+\alpha T^{2} + \beta T^{5}+ R_{K}\left( T/T_K\right) 
\end{equation}
Here, $R_{0}$ is the residual resistance due to disorder and impurities, $T^{2}$ is the e-e interaction term and $T^{5}$ is the contribution from the electron-phonon interaction. $ R_K(T/T_K) $ represents the contribution of magnetic ions to electrical resistivity. $ R_K(T/T_K) $ has a logarithmic trend at $ T \gg T_{K} $ and a saturating behavior at $ T \ll T_{K} $. 
 
To fit our experimental data we used the empirical form of $ R_K(T/T_K) $ presented by Costi \cite{Costi} and Goldhaber \cite{Goldhaber}, and later used by Lee \textit{et al.} in Ref. [\onlinecite{Gated-STO}] for undoped STO gated by an ionic gel electrolyte. They have demonstrated a Kondo effect, as a function of an applied electric field, which is the result of interaction between magnetic \textrm{Ti$^{3+}$} ions, unpaired and localized, with delocalized electrons that partially fill the \textrm{Ti} 3\textit{d} band. \par

\begin{equation}\label{fit-eq}
R_{K}\left(\dfrac{T}{T_{K}} \right)= R_{K}(t=0) \left(  \dfrac{T'^{2}_{K}}{T^{2}+T'^{2}_{K}}  \right)^{s}  
\end{equation}

Here, $T_{K}$ is defined as the temperature at which the Kondo resistivity is half of the value of resistivity at zero temperature, $ R_K(T_K)=R_K(T=0)/2 $ so that $ T'_{K}= T_{K}/\sqrt{2^{1/s}-1} $. $s$ is the effective spin of the magnetic scattering centers. We chose $ s=0.75 $ for both the $10^{-5}$ and $10^{-4}$ torr samples. For $10^{-3}$ torr sample, we first fit the model to high temperature part of the curve and extracted $ \alpha $, $ \beta $ and $ R_{0} $, and then used these values for fitting to low temperatures (below 10 K) and found $ T_{K} $. Table \ref{Rtable} summarizes the obtained values extracted from the fitting to equations \ref{R(T)-eq} and \ref{fit-eq} across the whole measured temperature range. 

\begin{table}[h!]
\begin{center} 
	\caption{Parameters obtained from fitting sheet resistance data of Fig. \ref{RvsT} to Eq. \ref{R(T)-eq} and \ref{fit-eq}. }
	\label{Rtable}
	\begin{tabular}{ | l | l | l | l|}
		\hline
		$P_{O_{2}}$ ($\mathrm{torr}$)  & $10^{-5}$ & $10^{-4}$ & $10^{-3}$ \\ \hline
		$s$ & 0.75 & 0.75 & 2.75 \\
		$R_{0}$ ($\mathrm{\Omega}$) & 2090 & 1540 & 7830  \\ 
		$\alpha$ ($\mathrm{\Omega/K^{2}}$) & 0.153 & 0.36 & 0.89\\ 
		$\beta$ ($\mathrm{\Omega/K^{5}}$) & $4.5 \times 10^{-11}$ & $3.59 \times 10^{-9}$ & $1.11 \times 10^{-8}$ \\ 
		$R_{K}$(T=0) ($\mathrm{\Omega}$) & 3400 & 26000 & 549060\\ 
		$T_{K}$ ($\mathrm{K}$) & 16.9 & 17.3 & 8.5 \\\hline					
\end{tabular}
\end{center}
\end{table}

From Table \ref{Rtable}, we see a gradual increase in the $ T_{K} $ values as the $P_{O_{2}}$ is increased from $10^{-5}$ torr to $10^{-4}$ torr. Because of the unusual peak in R-T curve for $10^{-3}$ torr sample, the value of $ T_{K} $ is not consistent with other two samples. \par
\par
\begin{figure}[h]
	\includegraphics[width=\linewidth, trim=0 2.2cm 0 1cm, clip]{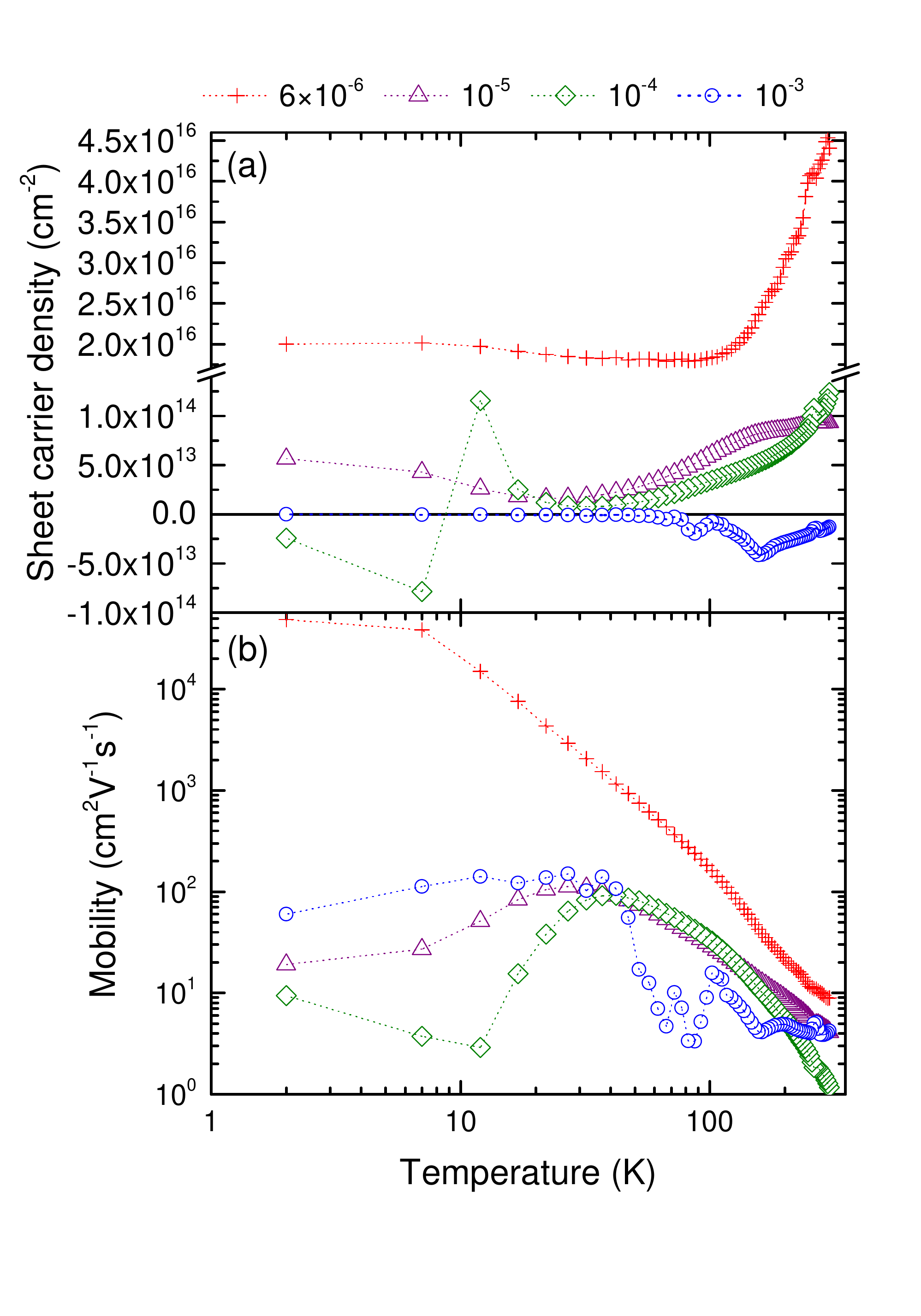}
	\caption{(Color online) Sheet carrier density, \textit{n$_s$} (a) and Hall mobility, $\mu$ (b) as a function of temperature.} \label{ns-mobility}
\end{figure}

The temperature dependences of sheet carrier density (\textit{n$_s$}) and mobility ($\mu$) are displayed in figures \ref{ns-mobility} (a) and (b), respectively. The positive values for \textit{n$_s$} indicate an electron-doped interface, and negative values indicate a hole-doped interface. For the $P_{O_{2}}=10^{-6}$ torr sample, the carrier concentration is two orders of magnitude larger than the amount predicted by the polar catastrophe ($3.2 \times 10^{14} $ \textrm{cm$^{-2}$}), if half an electron per unit cell is transfered to STO surface from LAO. The \textit{n$_s$} values for the lowest $P_{O_{2}}$ sample are so high that it resembles a bulk metallic sample. This implies that the oxygen vacancies are contributing to the numbers of donors or interdiffusion and creating a thick conducting region for the $P_{O_{2}}=10^{-6}$ torr sample.

The behavior of \textit{n$_s$} is rather complex and may involve two types of carriers freezing out at low temperatures for interfaces grown at $P_{O_{2}} 	\ge 10^{-4}$ torr. At higher temperatures, the $10^{-3}$ torr sample is predominantly doped with holes, which eventually freeze out at low temperatures; at 2 K, the carriers are electrons with a concentration of $1.9\times10^{11}$ \textrm{cm$^{-2}$}. The carrier type for the $10^{-4}$ torr sample also changes below 10 K. The two lowest oxygen-pressure grown samples are electron doped for all temperatures. For the $6\times10^{-6}$ torr sample, the mobility monotonically increases by three orders of magnitude as the sample is cooled down from room temperature, Fig. \ref{ns-mobility} (b). For other samples $\mu$ drops by further cooling below 40 K, perhaps due to localization. There are some variations in \textit{n$_s$} and $\mu$ at certain temperatures; those that happen at 150 K and 220 K are around the structural transitions of PAO in bulk. At room temperature, PrAlO$_3$ has a rhombohedral structure which transforms to an orthorhombic structure upon lowering the temperature to about 205--225 K, and to a monoclinic structure at about 151--175 K\cite{Pr-transitions,Pr-para}. The distinct drop of \textit{n$_s$} at 32 K has been previously seen for the NAO/STO interface\cite{Nd} and for the LAO/STO\cite{LAO-15uc} interface at around 20 K. \par
Surprisingly, the mobility values at low temperature are higher for the $10^{-3}$ torr sample than for the $10^{-4}$ torr sample, despite the latter having a higher resistance and lower \textit{n$_s$} at low temperatures. This could be in part due the difference in the number of defects or impurities in the two samples, for example, minimal oxygen vacancy defects in the $10^{-3}$ torr sample. 
\begin{figure} [h]
	\includegraphics[width=\linewidth, trim=0 1cm 0 0cm, clip]{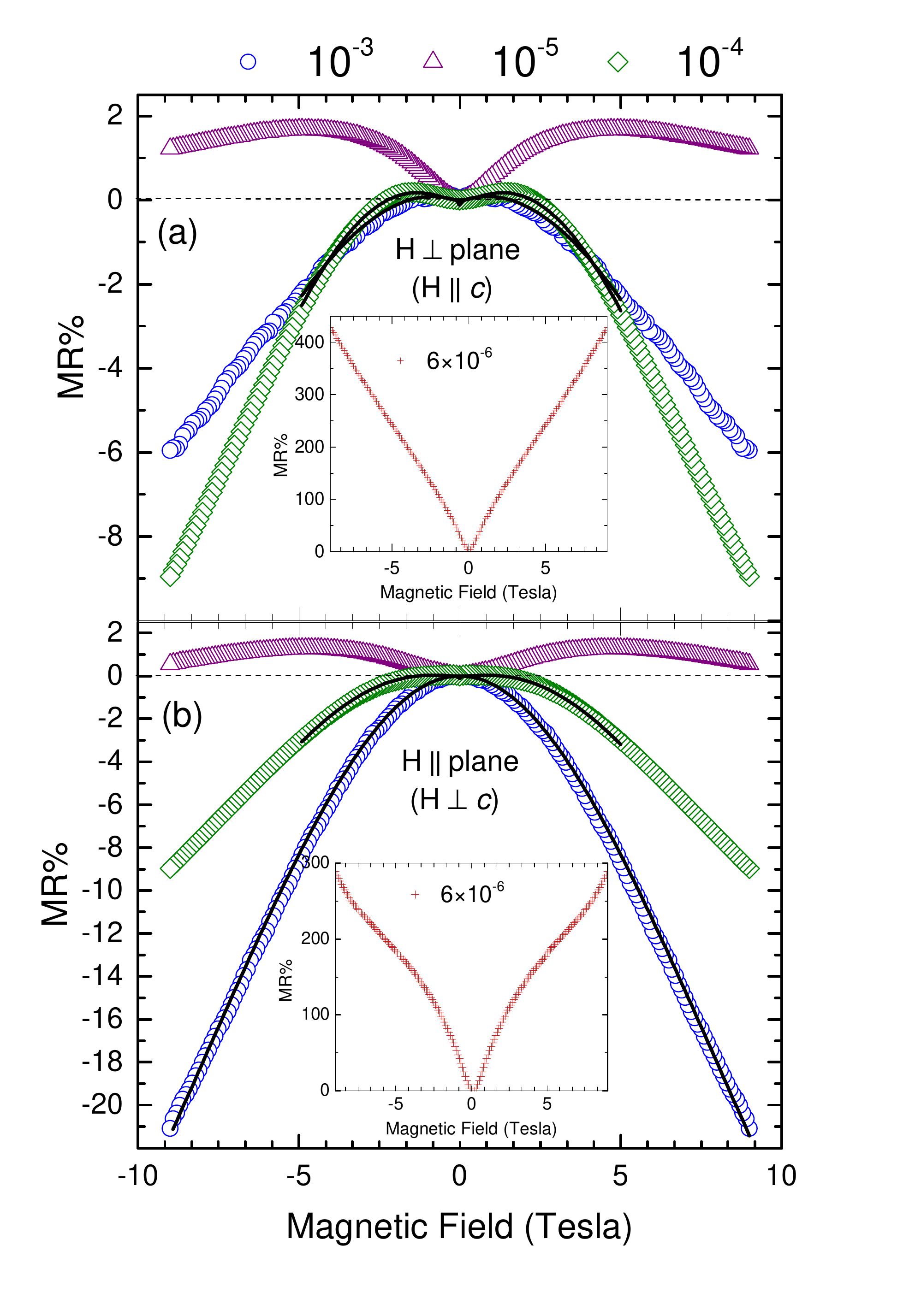}
	\caption{(Color online) MR of PAO/STO interfaces at 2 K. The magnetic field is applied (a) out of the plane and  (b) in the plane of the substrate. Insets show MR data for $6\times10^{-6}$ torr sample. Solid curves are fits according to Eq. \ref{KonWal}.} \label{MRs}
\end{figure}
\par
\subsection{Magnetoresistance and spin orbit interaction}
Figure \ref{MRs} (a) shows the magnetic field dependence of the magnetoresistance (MR) at 2 K. MR data are calculated as $\mathrm{MR=[\textit{R}\left( \textit{H}\right)-\textit{R}\left( 0\right) ]/\textit{R}(0)\times100\%}$. The measurement was done both: with the magnetic field \textit{H} perpendicular to the interface plane (out-of-plane MR), Fig. \ref{MRs} (a); and with the magnetic field \textit{H} parallel to the interface plane (in-plane MR), Fig. \ref{MRs} (b). \par
  
For the $6\times10^{-6}$ torr sample, MR values are positive for all measured fields and are immensely large (insets of Fig. \ref{MRs} (a) and (b)); the field dependence resembles some bulk metals. Very high positive MR has also recently been reported in degenerate semiconductive STO single crystals capped with ultra thin STO/LAO bilayers\cite{David2015}.\par
For higher $P_{O_{2}}$ samples, the MR data has a positive slope at low fields and a negative slope at high fields, resulting in a local maximum in the MR. These variations indicate that at least two different field-dependent scattering processes operate at the interface. The maximum of the MR shifts to a lower field as the oxygen pressure during growth increases. For the $10^{-3}$ torr sample the positive part of the MR is small.\par
A MR anisotropy investigation\cite{La-MR} for 26 u.c. LaAlO$_3$ on SrTiO$_3$  showed that the sign of the MR does not depend on the field direction for low $P_{O_{2}}$ samples, considered to be 3D samples (both the out of plane and in-plane MRs are positive for fields up to 9 T). However, for high $P_{O_{2}}$ LAO/STO the 2D samples, the sign of the MR changed from positive values in the out-of-plane geometry to negative values in the in-plane field orientation. For both 2D and 3D LAO/STO interfaces, the magnitude of the MR values are smaller when the field is in the plane of the interface. MR was observed to be negative for 12 uc of NAO/STO grown at $7.5\times10^{-3}$ torr\cite{Nd}. \par  

 We attribute the positive part of the MR to strong spin-orbit (SO) scattering and the negative part to the Kondo mechanism. The SO interaction has been shown to be very strong in some of the semiconductor heterostructures such as \textit{p}-type GaAs/Al$_x$Ga$_{1-x}$As\cite{GaAs}, and in topological insulators\cite{TI-review,TI-Taiwan}. The constructive interference between time-reversed waves of carriers in a disordered material leads to the weak localization (WL) effect\cite{WL}. Applying a magnetic field perpendicular to the plane of motion of carriers inhibits the WL and causes a negative MR at small magnetic fields. A strong SO interaction decreases the probability of carriers backscattering and counteracts weak localization. This effect is called weak antilocalization (WAL) and leads to a positive MR at small fields.\par
\begin{figure} [h]
  	\includegraphics[width=9 cm]{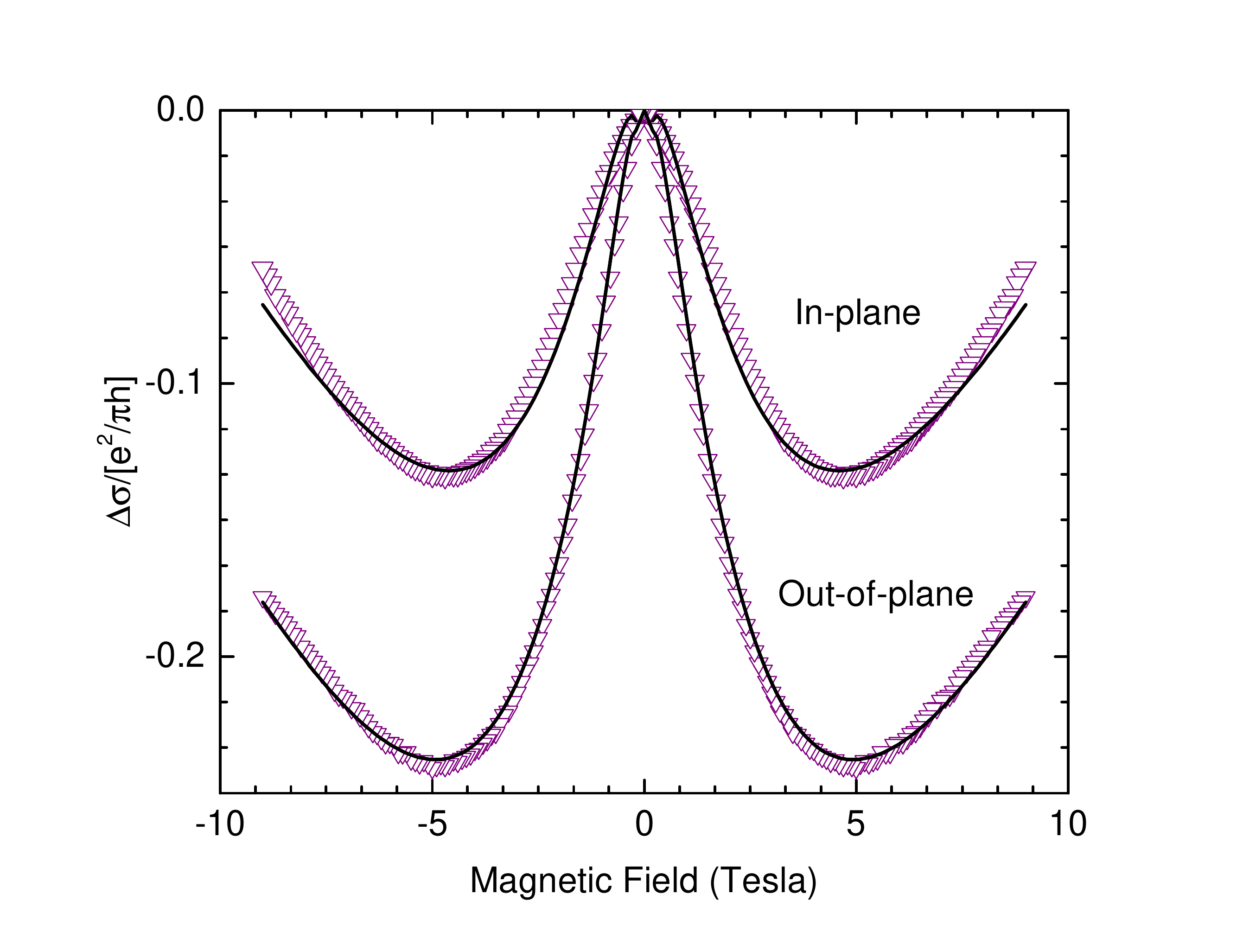}
  	\caption{Dependence of magneto-conductance, expressed in units of $ e^{2}/\pi h $, on magnetic field at 2 K for the $10^{-5}$ torr oxygen growth pressure sample for both field orientations. Black lines are fit to Eq. \ref{S-O for conductivity}.} \label{conductivity}
  \end{figure}

The conduction electrons confined in the vicinity of the polar RAO/STO interface experience a strong electric field perpendicular to the interface as the result of broken inversion symmetry. In the carriers' rest frame, this electric field appears as an internal magnetic field at the interface plane, perpendicular to their wave vector. As a result, the spins of electrons couple to the internal magnetic field which leads to a large Rashba SO interaction whose magnitude is tunable by the application of an external electric field\cite{LAO-Rashba}. We use the same analogy to fit the MR data for our samples. We fit the positive MR data at the $10^{-5}$ torr sample with the Hikami-Larkin-Nagaoka (HLN) theory\cite{Hikami}. HLN theory considers the effect of SO scattering, random magnetic impurity scattering, magnetic field, and inelastic collisions on the quantum backscattering interference,  and is expressed as a change in conductivity:

\begin{equation}\label{S-O for conductivity}
\begin{split}
\Delta \sigma(H) & = -\dfrac{e^{2}}{\pi h}\Bigg[ \frac{1}{2}\Psi\left( \dfrac{1}{2}+\frac{H_{\varphi}}{H}\right) -\ln\left( \frac{H_{\varphi}}{H}\right) \\
&-\Psi\left( \dfrac{1}{2}+\frac{H_{\varphi}+H_{SO}}{H}\right) +\ln\left( \frac{H_{\varphi}+H_{SO}}{H}\right) \\
&-\frac{1}{2}\Psi\left( \dfrac{1}{2}+\frac{H_{\varphi}+2H_{SO}}{H}\right) +
\frac{1}{2}\ln\left( \frac{H_{\varphi}+2H_{SO}}{H}\right) \Bigg]
\end{split}
\end{equation}

where $\sigma$ is the longitudinal conductivity and is calculated from the inversion of the measured resistivity matrix. $\Delta\sigma(H)=\sigma(H)-\sigma(0)$, $ \Psi(x) $ is the digamma function, $ e^{2}/\pi h $ is the universal value of conductance and $ H_{\varphi} = \hbar/4eD\tau_{\varphi} $. $ D $ and $ \tau_{\varphi} $ are the diffusion constant and phase coherence time (inelastic scattering time), respectively. $ H_{SO} = \hbar/4eD\tau_{SO} $ is the field at which the SO interaction is no longer effective and the positive WAL magnetoresistance becomes the negative WL magnetoresistance. $ \tau_{SO} $ is the SO scattering time. HLN theory was derived for materials in which the spin-splitting energy is proportional to \textit{$ k^{3} $}, which is the case in SrTiO$_3$\cite{K3SpinSplitting}. We treated both $H_{SO}$ and $ H_{\varphi}$ as fitting parameters and the good quality of the fit in a wide magnetic field range is shown for the $10^{-5}$ torr sample in Fig. \ref{conductivity}. While HLN theory is significant only in the diffusive regime, the fit, however, is shown for a wide magnetic field range. We obtained $ H_{\varphi}=0.33$ T  and $H_{SO}=1.25$ T for perpendicular, and $ H_{\varphi}=0.50 $ T  and $H_{SO}=1.25$ T for parallel magnetic field orientations, respectively, for the $10^{-5}$ torr sample.
 
As the $P_{O_{2}}$ is increased the positive part of the MR becomes smaller. The $10^{-4}$ torr sample has positive MR at magnetic fields near $ H=0 $. The maximum of its MR happens at 1.5 T; for the $10^{-3}$ torr sample, at 0.5 T. To fit the experimental data for samples grown at $P_{O_{2}}=10^{-4} $ and $10^{-3} $ torr to theory, we add the WAL positive MR, due to strong SO interaction, to the negative Kondo MR: 

\begin{equation}\label{KonWal}
	R\left(H \right)= R_{0}+R_{K}\left(\dfrac{H}{H_{1}} \right) + c \left[\Psi(\dfrac{1}{2}+\frac{H_{\varphi}}{H})-\ln(\frac{H_{\varphi}}{H})\right] 
\end{equation}

where $c$ is a constant and $ H_{1} $ is a magnetic field scale which is related to both $ T_{K} $ and the \textit{g}-factor of the impurity spin. $ R_{K}(H/H_{1}) $ is the zero temperature Kondo-magnetoresistance (taken at 2 K for our data), which is expressed in Ref. [\onlinecite{RevModPhys}], as:

\begin{equation}\label{R_k(H)}
	R_{K}\left(\dfrac{H}{H_{1}} \right)= R_{K}(H=0) \cos^2 \left(  \dfrac{\pi}{2}M(\dfrac{H}{H_{1}})\right) 
\end{equation}

 $ M(\dfrac{H}{H_{1}}) $ is the magnetization of a Kondo impurity at zero temperature and is given by: 
 
 \begin{widetext}
 	\begin{equation}\label{Mag_resis}
 		M(\dfrac{H}{H_{1}})=
 		\begin{cases}
 			\dfrac{1}{\sqrt{2\pi}}\displaystyle\sum_{i=0}^{\infty}(-\dfrac{1}{2})^i(i!)^{-1}(i+\dfrac{1}{2})^{(i-\dfrac{1}{2})}\mathrm{e}^{-(i+\dfrac{1}{2})} (\dfrac{H}{H_{1}})^{2i+1}       & \quad  H\leq\sqrt{2}H_{1} \\
 			1-(\pi)^{-3/2}\int\dfrac{dt}{t}\sin(\pi t) \mathrm{e}^{-t \ln(t/2e)} (\dfrac{H}{H_{1}}) 	\Gamma(t+1/2) & \quad H\geq\sqrt{2}H_{1}\\
 		\end{cases}
 	\end{equation}
 \end{widetext}
 
This interpretation was previously used by Lee \textit{et al.}\cite{Gated-STO} in electrolyte gated STO to explain the negative MR regime, and by Das \textit{et al.}\cite{hendi} in $ \delta $-doped  LaTiO$_{3}$/SrTiO$_{3} $ with LaCrO$_{3}$ to explain the small positive MR which is followed by a transition to a negative MR regime. The result of fitting to Eq. \ref{KonWal} is shown with black solid lines in Fig. \ref{MRs}. The WAL effect is meaningful for small fields, $ H < H_{SO} $, thus we only fit the data for small field range.  The fitting parameters are listed in table \ref{MRtable}.  

\begin{table}[h]
	\centering
	\caption{Parameters extracted from fitting the MR data in Fig. \ref{MRs} to Eq. \ref{KonWal}.}
	\label{MRtable}
	\begin{tabular}{|l|c|c|c||c|c|c|}
		\hline
		\multirow{2}{*}{$P_{O_{2}}$ (torr)} & \multicolumn{3}{c||}{Out-of-plane} & \multicolumn{3}{c|}{In-plane} \\
		& $H_{1}(T)$   & $H_{\varphi}(T)$   & c      & $H_{1}(T)$  & $H_{\varphi}(T)$  & c    \\ \hline
		$10^{-4}$                    & 11.4        & 0.46         & 800    & 11.6     & 0.52       & 800  \\ 
		$10^{-3}$                    & 15      & 0.15          & 30000   & 8.72     & --        & 0    \\ \hline
	\end{tabular}
\end{table}

\par
When the field is brought from being perpendicular to being parallel to the interface (Fig. \ref{MRs} (b)), there is no sign change in the MR values. For the in-plane geometry, for the $6\times10^{-6}$ torr sample, a cusp shape appears for $ H > $ 5 T , which implies the presence of an additional scattering mechanism that becomes active at higher fields. The in-plane MR values are smaller for most of the samples, but for the $10^{-3}$ torr sample they are larger. The positive part of the MR, which we assign to strong SO interaction, is very small for the $10^{-4} $ torr sample and it is negligible for the $10^{-3} $ torr sample for the in-plane orientation. For the high $P_{O_{2}}$ sample, in which carriers are just coming from electronic reconstruction and there are very few oxygen vacancies present, the in-plane data are smaller than corresponding out-of-plane data. This is consistent with the expected 2D nature of this interface.

\section{Conclusion}\label{Conclusion}
 In summary, we examined the behavior of the conductivity at the interface between PrAlO$_{3}$ and SrTiO$ _{3} $; we studied the electric and magnetic transport properties of interfaces of PAO/STO grown at oxygen pressures in the $P_{O_{2}}$ range of $6\times10^{-6}$ -- $1\times10^{-3}$ torr. For the $6\times10^{-6}$ torr sample the very small value of $\textit{R$_s$}$ at low temperature, large carrier concentration, high mobilities, and very high positive MR values indicate it is like a bulk metal, likely due to extensive oxygen defects and/or interdiffusion. As the interfaces are grown at higher oxygen pressures, resistivity behavior characteristic of nearly two-dimensional transport occurs; for $10^{-5}$ and $10^{-4}$ torr, high-temperature metallic behavior with increasing $\textit{R$_s$}$ is accompanied by an upturn at low temperature, consistent with Kondo scattering theory. Analysis of the $\textit{R$_s$}$ data gives Kondo temperature $ \sim $ 17 K for both the $10^{-5}$ and $10^{-4}$ torr samples. 
 The interface grown at $10^{-3} $ torr has the highest resistance with an usual peak at 57 K, and is mostly hole doped; most of its charge carriers freeze out at low temperatures, though the carrier mobility increases at those temperatures. 
 The MR values for the $10^{-4}$ and $10^{-3}$ torr samples were modeled with their positive part due to WAL because of a strong SO interaction and their negative part due to the Kondo effect. The variation of MR values suggests a strong SO interaction for the $10^{-5} $ torr sample with $ H_{SO} $ = 1.25 T in both field orientations. The positive part of the MR shrinks for the $10^{-4} $ and $10^{-3} $ torr samples. For these samples the MR is dominated by negative Kondo MR at high fields. The 2DEG has the lowest carrier density for the $10^{-3} $ torr sample, and comparison of MR values with theory suggest a high Kondo field scale $H_{1}(T)=$11 T for this sample for both field orientations.

\section*{Acknowledgment}
The authors acknowledge support from the Robert A. Welch Foundation Grant No. F-1191 and from the University of Texas at Austin College of Natural Sciences Freshman Research Initiative. 

\section*{References}


%

\end{document}